\documentclass[floatfix,aps,showpacs,superscriptaddress,twocolumn]{revtex4}
\usepackage{amsmath,amsfonts,amssymb}
\usepackage{graphicx}
\usepackage{dsfont}
\begin{document}
\title{Bogoliubov Hamiltonian as Derivative of Dirac Hamiltonian via Braid Relation}
\author{Bao-Xing Xie}
\email{xbx@mail.nankai.edu.cn}
\affiliation{Chern Institute of Mathematics, Nankai University,
Tianjin, 300071, China}
\author{Kang Xue}
\affiliation{Dept. of Physics, North East Normal
University, Changchun, 130024, China}
\author{Mo-Lin Ge}
\email{geml@nankai.edu.cn} \affiliation{Chern Institute of
Mathematics, Nankai University, Tianjin, 300071, China}
\begin{abstract}
   In this paper we discuss a new type of  4-dimensional representation
   of the braid group.
    The matrices of braid operations are constructed by
    $q$-deformation of Hamiltonians. One is the Dirac Hamiltonian for free
    electron with mass $m$, the other, which we find, 
	is related to the Bogoliubov Hamiltonian for quasiparticles 
	in $^3$He-B with the same free energy and mass being $m/2$.
	In the process, we choose the free $q$-deformation parameter as a special
    value in order to be consistent with the anyon description for
    fractional quantum Hall effect with $\nu=1/2$.
\end{abstract}
\vspace{0.3cm} \pacs{03.65.Fd,03.67.-a,67.30.-n}
\bigskip
 \maketitle

 \textit{Introduction and motivation.} The Bogoliubov Hamiltonian for quasiparticles in $^3$He-B is
compared to Dirac Hamiltonian for electrons, for example see Ref. \cite{Volovik}. It can be
expressed as (we have employed $\hbar=c=1$):
\begin{equation}
H_B=m(\vec{k})\beta+\vec{p}\cdot\vec{\alpha} \label{BH}
\end{equation}
where the mass $m(\vec{k})$ and momentum $\vec{p}$ are
\begin{equation}
    m(\vec{k})=\frac{\vec{k}^2}{2m}-\mu,~\vec{p}=\vec{k}\Delta_B/k_F,
\end{equation}
$\Delta_B$ is equilibrium order parameter, $k_F$ is Fermi momentum, and
$\beta$ and $\vec{\alpha}$ are $4\times4$ Dirac matrices:
\begin{equation}
~\beta=\left(
\begin{array}{cc}
    \mathds{1}_2&0\\
    0&-\mathds{1}_2
\end{array}
\right),~\vec{\alpha}=\left(
\begin{array}{cc}
    0&\vec{\sigma}\\
    \vec{\sigma}&0
\end{array}
\right)
~\textrm{($\vec{\sigma}$ : Pauli matrices).}
\end{equation}
It is similar to the Dirac Hamiltonian for the
electrons with mass $m$:
\begin{equation}
H=m\beta+\vec{p}\cdot\vec{\alpha} \label{DH}
\end{equation}
The difference is in  the mass term. The Bogoliubov Hamiltonian
essentially depends on $k$, and is not invariant under Lorentz group
while the Dirac equation is. In this paper, we would like to show
that the Bogoliubov Hamiltonian Eq. \eqref{BH} can be connected with
Dirac Hamiltonian Eq. \eqref{DH} through the braid relation (for braid group,
see, for example \cite{Jimbo,Kauffman}).

Our idea comes from the  Dirac game trying to describe that a spinor
needs to be rotated $4\pi$ to resume to the initial state. C. N. Yang
discussed this problem and showed an explanation using braid
group\cite{YangLetter}. Since Yang's comment is not published, we
firstly would like to review his result in the following. If one
defines $a$,$b$ as the following operations
    \begin{center}
             \includegraphics[width=0.1\textwidth]{a0a.eps}
             ~ ~ ~ ~
             \includegraphics[width=0.1\textwidth]{b0b.eps}
    \end{center}
then the braid group relation reads
\begin{equation}
    aba=bab
    \label{aba}
\end{equation}
that together with the additional rule for Dirac game
\begin{equation}
    abba=1
    \label{abba}
\end{equation}
imply that
\begin{equation}
    a^4=b^4=1.
    \label{a4}
\end{equation}
A diagrammatic illustration of Eq. \eqref{aba} and Eq. \eqref{abba}
is given by Fig. \ref{fig:aba} and Fig. \ref{fig:abba}.
\begin{figure}[h]
    \begin{minipage}[t]{0.45\linewidth}
        \centering
        \includegraphics[width=1.1\textwidth]{braidrelation1.eps}
        \caption{The braid  group relation}
        \label{fig:aba}
    \end{minipage}
    \hfill
    \begin{minipage}[t]{0.45\linewidth}
        \centering
        \includegraphics[width=\textwidth]{abba2.eps}
        \caption{The additional rule for Dirac game}
        \label{fig:abba}
    \end{minipage}
\end{figure}
The Dirac game is a model of spin-1/2  particle.
 In this paper, however, we take the operations 
 $a,b$ as the operations on anyons discussed in \cite{anyons} and they also
 satisfy the braid relation Eq. \eqref{aba}.
Instead of satisfying Eq. \eqref{a4},
for the simplest anyon model  they satisfy
\begin{equation}
	a^8=b^8=1,
	\label{ab_1}
\end{equation}
Which means a anyon needs to be rotated $8\pi$ to resume its initial state.
In this condition we look for what happens if we connect $a$ with
the Dirac Hamiltonian.
 We use the $q-$deformation of the normalized Dirac Hamiltonian
 in the form $a=\exp(i\frac{\theta}{2E}H)$, where $E$ is the energy,
 i.e. $H\Psi=E\Psi$ and $\theta$ is a constant and find $b$ in
 terms of Eq. \eqref{aba}. With the found $b$ we define the new Hamiltonian
 through $b=\exp(i\frac{\theta}{2}\mathcal{H})$.
 Finally we find that $\mathcal{H}$ is the
 Bogoliubov's Hamiltonian divided by its energy $\mathcal{E}$ (see
 below, Eq. \eqref{H}).

\textit{Derivation.} For a free electron the normalized  Dirac
Hamiltonian is given in standard notation by \cite{Shankar&Mathur}
\begin{eqnarray}
 H_D =\frac{1}{E}(m\beta+\vec{p}\cdot\vec{\alpha})
\end{eqnarray}
with the normalization $E=\sqrt{p^2+m^2},~p=|\vec{p}|$ and
$H_D^2=1$, as introduced in Ref. \cite{Shankar&Mathur}. We introduce
the $q$-deformation of $H_D$ in terms of
$R_1(\theta)=e^{i\frac{\theta}{2}H_D}$. Then we can get a
q-deformation of another Hamiltonian $\mathcal{H}$, i.e.
$R_2(\theta)=e^{i\frac{\theta}{2}\mathcal{H}}$ based on
Eq. \eqref{aba}. As we show below, $R_1$ and $R_2$ satisfy the  braid
relation:
\begin{equation}
        R_1R_2R_1=R_2R_1R_2.
    \label{origin_braid}
\end{equation}
We shall look for a new type of solution for $R_1$ and $R_2$
that can not be written as $R\otimes \mathds{1}$ and
$\mathds{1}\otimes R$ where $R$ and $\mathds{1}$ are
2-D matrix and identity, respectively.

After preforming a unitary transformation on both $R_1$ and $R_2$, they of
course still satisfy the same relation. The unitary
matrix
\begin{equation}
    V=\frac{1}{\sqrt{2E}}
    \left(
    \begin{array}{cccc}
        p_+u_-^{-1} &-p_3u_-^{-1} & 0 & u_- \\
        -p_+u_+^{-1}&p_3u_+^{-1}  & 0 & u_+ \\
        p_3u_-^{-1} &p_-u_-^{-1}  & u_- & 0 \\
        -p_3u_+^{-1}&-p_-u_+^{-1} & u_+ & 0
    \end{array}
    \right)
    \label{1}
\end{equation}
with $p_{\pm}=p_1\pm ip_2,~u_{\pm}=\sqrt{E\pm m}$, transforms the
$H_D$ diagonalized:
\begin{equation}
    VH_DV^{\dagger}=\left(
    \begin{array}{cc}
        \sigma_3 & 0 \\
        0 & \sigma_3
    \end{array}
    \right).
    \label{2}
\end{equation}
It follows that in terms of such  transformation
$R_1(\theta)$ becomes diagonal as
well and is  composed of block diagonal form:
\begin{eqnarray}
    R_1'(\theta)=Ve^{i\frac{\theta}{2}H_D}V^{\dagger}&=&\left(
    \begin{array}{cccc}
        e^{i\frac{\theta}{2}} & 0 & 0 & 0 \\
        0 & e^{-i\frac{\theta}{2}} &0 &0 \\
        0&0&e^{i\frac{\theta}{2}} &0 \\
        0&0&0&e^{-i\frac{\theta}{2}}
    \end{array}
    \right)
    \label{3}
    \\\nonumber
    &=&\left(
    \begin{array}{cc}
        a(\theta) &0 \\
        0&a(\theta)
    \end{array}
    \right),
\end{eqnarray}
where $a(\theta)=\left(
\begin{array}{cc}
    e^{i\frac{\theta}{2}}&0\\
    0&e^{-i\frac{\theta}{2}}
\end{array}
\right)
=e^{i\frac{\theta}{2}}\left(
\begin{array}{cc}
    1 &0 \\
    0&e^{-i\theta}
\end{array}
\right) $.

Until now, $\theta$ can be a free parameter. One
particular choice is related to the anyon description for
the fractional
quantum Hall effect (FQHE) with $\nu=1/2$ \cite{anyons}, let
$\theta=-\frac{\pi}{2}$, then
\begin{equation}
    a=a(\theta=-\frac{\pi}{2})=e^{-i\pi/4}\left(
    \begin{array}{cc}
        1&0\\
        0&i
    \end{array}
    \right).
    \label{a}
\end{equation}
With the choice of $\theta=-\frac{\pi}{2}$ we can find
a matrix $b$ satisfying the braid group relation Eq. \eqref{aba}.
A typical solution is given by \cite{preskill}
\begin{equation}
    b=\frac{1}{\sqrt{2}}\left(
    \begin{array}{cc}
        1&i\\
        i&1
    \end{array}
    \right).
    \label{b}
\end{equation}
In fact, both matrices $a$ and $b$ emergence exactly at the anyon
description for FQHE with $\nu=1/2$ \cite{anyons} and satisfy $a^8=b^8=1$.

Under the action of the unitary transformation $V$, we expect
$R_2'(\theta)=Ve^{i\frac{\theta}{2}\mathcal{H}}V^{\dagger}$ take a
block diagonal representation.  Therefore, the matrix
\begin{equation}
    R_2'=\left(
    \begin{array}{cc}
        b&0\\
        0&b
    \end{array}
    \right)
    =\frac{1}{\sqrt{2}}\left(
    \begin{array}{cccc}
        1&i&0&0\\
        i&1&0&0\\
        0&0&1&i\\
        0&0&i&1
    \end{array}
    \right)
    \label{8}
\end{equation}
naturally satisfies the relation
\begin{equation}
        R_1'R_2'R_1'=R_2'R_1'R_2' 
    \label{9}
\end{equation}
Utilizing the same transformation $V$ to switch back, the calculation gives
\begin{eqnarray}
    R_2=V^{\dagger}R_2'V&=&\frac{1}{\sqrt{2}}(1-i\frac{1}{\frac{Ep}{m}}(\frac{p^2}{m}\beta-\vec{p}\cdot\vec{\alpha})) \nonumber \\
	&=&\frac{1}{\sqrt{2}}(1-i\mathcal{H}).
    \label{R2_t}
\end{eqnarray}
Where
\begin{equation}
    \mathcal{H}=\frac{1}{\frac{Ep}{m}}(\frac{p^2}{m}\beta-\vec{p}\cdot\vec{\alpha}).
\end{equation}
Noting that $\mathcal{H}^2=1$, we can recast $R_2$ to
\begin{equation}
	R_2=\frac{1}{\sqrt{2}}(1-i\mathcal{H})=e^{-i\frac{\pi}{4}\mathcal{H}}=
	e^{i\frac{\theta}{2}\mathcal{H}}|_{\theta=-\frac{\pi}{2}}.
    \label{R2_o}
\end{equation}
Therefore $\mathcal{H}$ is just the Hamiltonian we are looking for.
Finally, we get
\begin{equation}
    \mathcal{H}=\frac{1}{\frac{Ep}{m}}(\frac{p^2}{m}\beta-\vec{p}\cdot\vec{\alpha})
    =\frac{1}{\mathcal{E}}(\frac{p^2}{m}\beta-\vec{p}\cdot\vec{\alpha}),~\mathcal{E}=\frac{p}{m}E.
    \label{H}
\end{equation}
%
The obtained Hamiltonian $\mathcal{H}$ has a natural explanation:
It is the Bogoliubov
Hamiltonian for quasiparticles in $^3$He-B with the free energy
$\mu=0$ and mass being $m/2$ and $-\vec{p}$.

\textit{Conclusion and discussion.}
We have shown that the  Dirac Hamiltonian Eq. \eqref{DH} and
Bogoliubov Hamiltonian Eq. \eqref{BH} can be  related with
each other by the braid relation Eq. \eqref{aba}.
It is known that the braid matrices $R_1'$ and
$R_2'$ given by Eq. \eqref{3} and Eq. \eqref{8} are related to the
anyon description for FQHE with $\nu=1/2$ \cite{anyons}.
If there is
evidence of physical consequence of associative Hamiltonian
$\mathcal{H}$ based on $H_D$ through braid relation, we may expect
that Dirac particle could be decomposed to anyons on the basis of
the braid relation. 
%

\begin{figure}[ht]
     \includegraphics{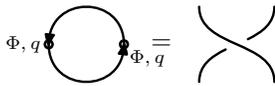}
     \caption{Interchange of two flux-tube-particle composites by a rotation}
     \label{fig:flux}
 \end{figure}

 In our derivation, We have decomposed the $4\times4$
 representation into two $2\times2$ representations. Therefore  if we
 denote in 2 dimension the interchange of
 two flux-tube-particle composites by a rotation  \cite{Wilczek}
 as Fig. \ref{fig:flux},
 then $a,b$ which defined as before must satisfy the braid
 relation Eq. \eqref{aba} for non-abelian models.
 This was discussed by Preskill \cite{preskill} and in
 path-integral formulation by Wu in \cite{Wu}.
     It is well known that the
braid relation Eq. \eqref{aba} is an asymptotic limit of  Yang-Baxter
equation (YBE) \cite{Jones,Yang&Ge,Chen&Xue&Ge} and in quantum field
theory YBE describes the scattering of particles in ($1+1$)
space-time \cite{Yang}. From the aspect of the related  YBE
describes the scattering of two ``particles'', which may be anyons,
we  might be brave to imaging a possible decomposition of a Dirac particle.
However, The nature of our result is still not clear.

  {\bf Acknowledgements}
  The authors would like to thank Prof. C. N. Yang for enlighten
  discussions and J. L. Chen and S. W. Hu for their
  helpful discussions. This work was supported in part by NSF of
  China (Grants No. 10575053) and The Cooperation Research Fund for LiuHui 
  Center for Applied Mathematics,
  Nankai University and Tianjin University.

\end{document}